\documentclass[12pt,aps,prb,preprint,showpacs,showkeys]{revtex4}  

\usepackage{amsmath}    
\usepackage{amsfonts}   
\usepackage{amssymb}
\usepackage{graphicx}   
\usepackage{subfigure}

\begin{document}
\baselineskip = 7.5mm \topsep=1mm
\begin{center}{\LARGE\bf Histogram Monte Carlo Simulation of the Geometrically Frustrated XY Antiferromagnet with Biquadratic Exchange}
\vspace{5 mm}\\ M.\v{Z}ukovi\v{c}$^{\rm a*}$, T.Idogaki$^{\rm a}$ and K.Takeda$^{\rm a,b}$ \vspace{5
mm}\\
\end{center}
\noindent $^{\rm a}$ Department of Applied Quantum Physics, Kyushu University
\newline
\noindent $^{\rm b}$ Institute of Environmental Systems, Kyushu University \vspace{3 mm}
\newline
\noindent {\bf{Abstract.}} Histogram Monte Carlo simulation is used to investigate effects of
biquadratic exchange $J_{2}$ on phase transitions of a 3D classical XY antiferromagnet with frustration
induced by the antiferromagnetic exchange $J_{1}$ and the stacked triangular lattice geometry. The
biquadratic exchange is considered negative (antiferroquadrupolar) within the triangular planes and
positive (ferroquadrupolar) between the planes. The phase diagram obtained features a variety of
interesting phenomena arising from the presence of both the biquadratic exchange and frustration. In a
strong biquadratic exchange limit ($|J_{1}|/|J_{2}| \leq 0.25$), the antiferroquadrupolar phase
transition which is of second order is followed by the antiferromagnetic one which can be either first
or second order. The separate antiferroquadrupolar and antiferromagnetic second-order transitions are
found to belong to the chiral XY and Ising universality classes, respectively. If the biquadratic
exchange is reduced both transitions are found to be first order and occur simultaneously in a wide
region of $|J_{1}|/|J_{2}|$. However, if $|J_{2}| \rightarrow 0$ the transition changes to the
second-order one with the chiral universality class critical behavior. \vspace{10 mm}
\\ $PACS\ codes$: 75.10.Hk; 75.30.Kz; 75.40.Cx; 75.40.Mg.
\newline
$Keywords$: XY model; Frustrated antiferromagnet; Biquadratic exchange; Phase transition; Quadrupole
ordering; Histogram Monte Carlo simulation;\\ \vspace{10 mm} \\ $*$Corresponding author.
\newline
Permanent address: Institute of Environmental Systems, Faculty of Engineering, Kyushu University,
Fukuoka 812-8581, Japan
\\ Tel.: +81-92-642-3811; Fax: +81-92-633-6958
\\ E-mail: milantap@mbox.nc.kyushu-u.ac.jp
\newpage
\noindent {\bf\Large{I. Introduction}} \vspace{3mm}
\newline
\indent The problem of biquadratic or generally higher-order exchange interactions in systems with
Heisenberg symmetry has been addressed in several mean field approximation (MFA) studies
\cite{sivar72}-\cite{chen-levy1}, by high-temperature series expansion (HTSE) calculations
\cite{chen-levy2}, as well as within a framework of some other approximative schemes
\cite{micnas,chaddha99}. It has been shown that such interactions can induce various interesting
properties such as tricritical and triple points, quadrupole ordering, separate dipole and quadrupole
phase transitions etc. Much less attention, however, has been paid to this problem on systems with XY
spin symmetry. Chen $et.al$ \cite{chen-etal1,chen-etal2} calculated transition temperatures and the
susceptibility critical indices for an XY ferromagnet with biquadratic exchange on cubic lattices by the
HTSE method for limited region of $J_{1}/J_{2}$. However, the rigorous proof of the existence of dipole
long-range order (DLRO), corresponding to the ferromagnetic directional arrangement of spins, and
quadrupole long-range order (QLRO), representing an axially ordered state in which spins can point in
either direction along the axis of ordering, at finite temperature on the classical bilinear-biquadratic
exchange model has only recently been provided independently by Tanaka and Idogaki \cite{tanaka}, and
Campbell and Chayes \cite{campbell}. Very recently we have considered the XY model with the
bilinear-biquadratic exchange Hamiltonian on a simple cubic \cite{nagata} and hexagonal (stacked
triangular) \cite{zukovic} lattices, and performed a finite-size scaling (FSS) analysis in order to
investigate critical properties of the considered systems via Standard Monte Carlo (SMC) and Histogram
Monte Carlo (HMC) simulations.
\newline
\indent So far, however, to our best knowledge there has been no investigation of the effect of the
biquadratic exchange on an XY model with frustrated/competing exchange interaction. In this paper we
present systematic investigations of the role of the biquadratic exchange in phase transitions of the
geometrically frustrated XY antiferromagnet on stacked triangular lattice (STL). This model has been
argued to possess some unique properties such as novel chiral universality class critical behavior
\cite{kawamura1,kawamura2}, but many more remarkable features have been observed when the effects of
external magnetic field \cite{plumer1} and next-nearest neighbors \cite{diep} were considered. In the
present work, the effect of the biquadratic exchange is also found to bring about variety of interesting
phenomena, such as regions of first order transitions, separate magnetic and quadrupolar ordering,
transitions of different universality classes, etc. \vspace{6mm}
\newline
\noindent {\bf\Large{II. Model and computation details}}\vspace{3mm}
\newline
\indent We consider the XY model, described by the Hamiltonian
\begin{equation}
H=-J_{1}\sum_{\langle i,j \rangle}\mbox{\boldmath $S$}_{i} \cdot \mbox{\boldmath $S$}_{j}
-J_{2}^{\perp}\sum_{\langle i,k \rangle}(\mbox{\boldmath $S$}_{i} \cdot \mbox{\boldmath $S$}_{k})^{2}
-J_{2}^{\parallel}\sum_{\langle i,l \rangle}(\mbox{\boldmath $S$}_{i} \cdot \mbox{\boldmath
$S$}_{l})^{2}\ ,
\end{equation}
where $\mbox{\boldmath $S$}_{i} = (S_{i}^{x},S_{i}^{y})$ is a two-dimensional unit vector at the $i$th
lattice site and the sums $\langle i,j \rangle$, $\langle i,k \rangle$ and $\langle i,l \rangle$ run
over all nearest neighbors (NN), NN in the $xy$-plane, and NN in the stacking $z$-axis direction,
respectively. We consider the bilinear exchange interaction $J_{1}<0$, the biquadratic intra-plane and
inter-plane exchange interactions $J_{2}^{\perp}<0$ and $J_{2}^{\parallel}>0$, respectively, with
$|J_{2}^{\perp}|=|J_{2}^{\parallel}|=|J_{2}|$.
\newline
\indent Assuming periodic boundary condition, spin systems of the
linear lattice sizes $L$ = 12, 18, 24 and 30 are first used in SMC
simulations. For a fixed value of the exchange ratio
$|J_{1}|/|J_{2}|$, we start the simulation process at low (high)
temperatures from an antiferromagnetic/random (random) initial
configuration and gradually raise (lower) temperature. These
heating-cooling loops serve to check possible hysteresis,
accompanying first-order transitions. As we move in
$(|J_{1}|/|J_{2}|,k_{B}T/|J_{2}|)$ space, we use the last spin
configuration as an input for calculation at the next point. We
sweep through the spins in sequence and updating follows a
Metropolis dynamics. In the updating process, the new direction of
spin in the spin flip is selected completely at random, without
any limitations by a maximum angle of spin rotation or allowed
discrete set of resulting angle values. Thermal averages at this
stage are calculated using at most $1\times10^{5}$ Monte Carlo
steps per spin (MCS/s) after thermalizing over another
$0.5\times10^{5}$ MCS/s. We calculate the system internal energy
$E$ and some other physical quantities defined as follows: the
specific heat per site $c$
\begin{equation}
\label{eq.c}c=\frac{(\langle E^{2} \rangle - \langle E \rangle^{2})}{Nk_{B}T^{2}}\ ,
\end{equation}
the dipole LRO (DLRO) parameter $m$,
\begin{equation}
\label{eq.m1}m=\frac{\langle M \rangle}{N}=\frac{1}{N}\left\langle\sqrt{6\sum_{\alpha=1}^{6}
\mbox{\boldmath $M$}_{\alpha}^{2}}\ \right\rangle\ ,
\end{equation}
where $\mbox{\boldmath $M$}_{\alpha}$ is the $\alpha$th sublattice-magnetization vector (note that the
present model has six equivalent magnetic sublattices), given by
\begin{equation}
\label{eq.m2}\mbox{\boldmath $M$}_{\alpha}=\left(\sum_{i}S_{\alpha i}^{x},\ \sum_{i}S_{\alpha
i}^{y}\right)\ ,
\end{equation}
the quadrupole LRO (QLRO) parameter $q$,
\begin{equation}
\label{eq.q1}q=\frac{\langle Q \rangle}{N}=\frac{1}{N}\left\langle\sqrt{6\sum_{\alpha=1}^{6}
\mbox{\boldmath $Q$}_{\alpha}^{2}}\ \right\rangle\ ,
\end{equation}
where
\begin{equation}
\label{eq.q2}\mbox{\boldmath $Q$}_{\alpha}=\left(\sum_{i}\left(\left(S_{\alpha
i}^{x}\right)^{2}-\left(S_{\alpha i}^{y}\right)^{2}\right),\ \sum_{i}2S_{\alpha i}^{x}S_{\alpha
i}^{y}\right)\ ,
\end{equation}
the chiral LRO (CHLRO) parameter $\kappa$,
\begin{equation}
\label{eq.ch1}\kappa=\frac{\sqrt{\langle K^{2} \rangle}}{N}=\frac{1}{N}\sqrt{\left\langle \left(\sum_{p}
\kappa_{p}\right)^{2} \right\rangle}\ ,
\end{equation}
where the summation runs over all upward triangles on the triangular layer and $\kappa_{p}$ represents a
local chirality at each elementary triangular plaquette, defined by
\begin{equation}
\label{chirality} \kappa_{p} = \frac{2}{3\sqrt{3}}\sum_{\langle i,j \rangle}^{p}[\mbox{\boldmath
$S$}_{i} \times \mbox{\boldmath $S$}_{j}]_{z}\ =\
\frac{2}{3\sqrt{3}}[\sin(\varphi_{2}-\varphi_{1})+\sin(\varphi_{3}-\varphi_{2})+\sin(\varphi_{1}-\varphi_{3})]
\ ,
\end{equation}
where the summation runs over the three directed bonds surrounding each plaquette, $p$, and
$\varphi_{i}$ represents the $i$th spin angle. $\kappa_{p}$ is an Ising-like quantity representing the
sign of rotation of the spins along the three sides of each plaquette. Further, the following quantities
which are functions of the parameter $O$ ($=\ M,\ Q,\ K$) are defined: the susceptibility per site
$\chi_{O}$
\begin{equation}
\label{eq.chi}\chi_{O} = \frac{(\langle O^{2} \rangle - \langle O \rangle^{2})}{Nk_{B}T}\ ,
\end{equation}
the logarithmic derivatives of $\langle O \rangle$ and $\langle O^{2} \rangle$ with respect to
$\beta=1/k_{B}T$
\begin{equation}
\label{eq.D1}D_{1O} = \frac{\partial}{\partial \beta}\ln\langle O \rangle = \frac{\langle OE
\rangle}{\langle O \rangle}- \langle E \rangle\ ,
\end{equation}
\begin{equation}
\label{eq.D2}D_{2O} = \frac{\partial}{\partial \beta}\ln\langle O^{2} \rangle = \frac{\langle O^{2} E
\rangle}{\langle O^{2} \rangle}- \langle E \rangle\ ,
\end{equation}
the fourth-order long-range order cumulant $U$ (Binder parameter)
\begin{equation}
\label{eq.U}U = 1-\frac{\langle O^{4}\rangle}{3\langle O^{2}\rangle^{2}} \ ,
\end{equation}
and the fourth-order energy cumulant $V$
\begin{equation}
\label{eq.V}V = 1-\frac{\langle E^{4}\rangle}{3\langle E^{2}\rangle^{2}}\ .
\end{equation}
The above quantities are useful for localization of a transition as well as for determination of its
nature. For example, first-order transitions usually manifest themselves by discontinuities in the order
parameter and energy, and hysteresis when cooling and heating. If transition is second order, it can be
localized approximately by the $\chi_{O}$ peak position or more precisely by the intersection of the
fourth-order LRO (or energy) cumulants curves for different $L$.
\newline
\indent In order to increase precision and reliability of the
obtained information, as well as to retrieve some additional
information which could not be extracted from the SMC
calculations, we further perform HMC calculations, developed by
Ferrenberg and Swendsen \cite{ferr-swen1,ferr-swen2}, at the
estimated transition temperatures for each lattice size. Here,
$2\times10^{6}$ MCS/s are used for calculating averages after
discarding another $1\times10^{6}$ MCS/s for thermalization. We
calculate the energy histogram $P(E)$, the order parameters
histograms $P(O)$\ $(O\ =\ M,\ Q,\ K)$, as well as the physical
quantities (\ref{eq.c})-(\ref{eq.V}). Using data from the
histograms, one can calculate physical quantities at neighboring
temperatures, and thus determine the values of extrema of various
quantities and their locations with high precision for each
lattice size. In such a way we can obtain quality data for FSS
analysis which determines the order of the transition and, in the
case of a second-order transition, it also allows us to extract
critical indices. For example, the energy cumulant $V$ exhibits a
minimum near critical temperature $T_{c}$, which achieves the
value $V^{*}=\frac{2}{3}$ in the limit $L\rightarrow \infty$ for a
second-order transition, while $V^{*}<\frac{2}{3}$ is expected for
a first-order transition \cite{ferr-swen1,ferr-swen2}.
Temperature-dependences of a variety of thermodynamic quantities
display extrema at the $L$-dependent transition temperatures,
which at a second-order transition are known to scale with a
lattice size as, for example:
\begin{equation}
\label{eq.scalchi}\chi_{O,max}(L) \propto L^{\gamma_{O}/\nu_{O}}\ ,
\end{equation}
\begin{equation}
\label{eq.scalD1}D_{1O,max}(L) \propto L^{1/\nu_{O}}\ ,
\end{equation}
\begin{equation}
\label{eq.scalD2}D_{2O,max}(L) \propto L^{1/\nu_{O}}\ ,
\end{equation}
\noindent where $\nu_{O}$ and $\gamma_{O}$ represent the correlation length and susceptibility critical
indices, respectively. In the case of a first-order transition (except for the order parameters), they
display a volume-dependent scaling, $\propto L^{3}$. The simulations were performed on the vector
supercomputer FUJITSU VPP700/56. \vspace{6mm}
\newline
\noindent {\bf\Large{III. Chirality on frustrated quadrupoles}}\vspace{3mm}
\newline
\indent It has been known for some time that the frustrated spin system on triangular lattice possesses
the chirality $\kappa$ as defined in Eqns.(\ref{eq.ch1},\ref{chirality}) \cite{miyashita}. Due to the
chirality the system has two-fold degeneracy of the ground state ($\kappa=+1$ and $\kappa=-1$),
resulting in the structure with spins arranged on plaquettes with turn angles $+120^{\circ}$ and
$-120^{\circ}$, respectively (Fig.1(a)). A minimum energy condition is realized by an arrangement in
which the $+$ and $-$ plaquettes alternate, producing long-range chiral order at low temperatures. Such
a system has been argued to belong to a nonstandard universality class linked to the two-fold chiral
degeneracy inherent to the $120^{\circ}$ ordered spin structure \cite{kawamura1,kawamura2}, the critical
behavior of which is characterized by critical indices, different from those for non-frustrated systems
with the same spin symmetry. Since the present Hamiltonian includes both bilinear and biquadratic terms,
let us take a closer look at the opposite side of the exchange ratio spectrum and investigate critical
behavior of the system with only biquadratic exchange interaction, i.e. the case of $J_{1}=0$. If
$J_{2}^{\perp}<0$ (the sign of $J_{2}^{\parallel}$ is irrelevant in the present consideration) the
quadrupolar system is frustrated due to the triangular lattice geometry, resulting in a non-collinear
ground state. The non-collinear ground state arrangement resembles the $120^{\circ}$ structure of the
antiferromagnetic system, however, here, the spins can point in either direction within the given axis
(for illustration see the snapshots in Fig.12). As far as the chirality $\kappa$ is concerned, such a
system has four-fold degeneracy in the ground state of each plaquette ($\kappa_{p}=\pm 1,\pm
\frac{1}{3}$), resulting in the structure with four possible turn angles between two neighboring spins
$\pm 120^{\circ}$, $\pm 60^{\circ}$. However, there is no energetically favorable arrangement among the
four kinds of plaquettes and, hence, the plaquettes do not order even at low temperatures. Nevertheless,
even for such a system we can define the quantity analogous to the chirality of the antiferromagnetic
system (let us call it the quadrupolar chirality) if we consider instead of spins their axes and turn
angles between the axes, which are again $\pm 120^{\circ}$. If we define the local quadrupolar chirality
as
\begin{equation}
\label{q-chirality} \kappa_{p}^{q} =
\frac{2}{3\sqrt{3}}[\sin2(\varphi_{2}-\varphi_{1})+\sin2(\varphi_{3}-\varphi_{2})+\sin2(\varphi_{1}-\varphi_{3})]
\ ,
\end{equation}
and the quadrupolar chirality LRO parameter (QCHLRO) $\kappa^{q}$ as
\begin{equation}
\label{eq.q-ch1}\kappa^{q}=\frac{\sqrt{\langle (K^{q})^{2} \rangle}}{N}=\frac{1}{N}\sqrt{\left\langle
\left(\sum_{p} \kappa_{p}^{q}\right)^{2} \right\rangle}\ ,
\end{equation}
concerning such defined quadrupolar chirality, the system will have two-fold degeneracy of the ground
state ($\kappa^{q}=-1$ and $\kappa^{q}=+1$, corresponding to turn angles $+120^{\circ}$ and
$-120^{\circ}$, respectively (Fig.1(b))), and the situation will much resemble the one for the
antiferromagnetic system with the chirality $\kappa$. Furthermore, in analogy with the chirality
$\kappa$ which is believed to order along with spins, here, the quadrupolar chirality $\kappa^{q}$ is
expected to show LRO simultaneously with quadrupoles. \vspace{6mm}
\newline
\noindent {\bf\Large{IV. FSS analysis and phase diagram}}\vspace{3mm}
\newline
\indent We first consider the case of $J_{2}=0$. To determine the
order of the transition we analyze the scaling behavior of the
minimal value of the energy cumulant $V$ at the transition
temperature. As shown in Fig.2, $V$ tends to the value of 2/3, as
expected for a second-order transition, and the slope 2.39 means
that $V$ is not volume dependent. Also, observing the energy and
LRO parameters distribution histograms (not shown), no bimodal
distribution, which would signal a first-order transition, is
found. Hence, both spin and chirality ordering transitions seem to
be clearly of second order. The transition temperature, calculated
from the intersection of the Binder parameter curves for different
$L$, is estimated to $k_{B}T_{c}/|J_{1}|=1.4580 \pm 0.0005$, in
agreement with the values quoted in Refs. \cite{kawamura1,diep}.
The chirality transition temperature
$k_{B}T_{c}^{\kappa}/|J_{1}|=1.4590 \pm 0.0013$, similarly as in
Ref. \cite{kawamura1}, seems to be slightly higher than the spin
ordering temperature but the two values cannot be distinguished
beyond the error bar and, hence, we assume they are the same. The
spin and chirality critical indices calculated from the scaling
relations (\ref{eq.scalchi})-(\ref{eq.scalD2}) take the following
values: $\nu_{M}=0.52\pm 0.03$, $\gamma_{M}=1.08 \pm 0.08 $ and
$\nu_{K} = 0.55 \pm 0.01$, $\gamma_{K} = 0.81 \pm 0.03$
\footnote{The errors for $\nu_{O}$ and $\gamma_{O}$ are calculated
from standard errors of the respective slopes $b$ in the linear
regression $y=a+bx$.}, respectively (Figs.3,4). Also the values of
the critical indices are in fair agreement with the two previous
studies \cite{kawamura1,diep}, however, as far as the universality
class is concerned the situation here is not so straightforward
and will be discussed later.
\newline
\indent The order of the transitions changes, however, when even a comparatively weak biquadratic
exchange interaction is introduced. Although it is very hard to observe the typical first-order behavior
for small values of $|J_{2}|$, if the lattice sizes are taken sufficiently large the signs of the
discontinuous transition show up. This is seen in Fig.5 in which the bimodal (double-peak) energy
distribution becomes clearly recognizable if $L \geq 30$, for the case of $|J_{2}|/|J_{1}|=\frac{1}{5}$.
As $|J_{2}|$ is increased, the first-order features of the transition are becoming more and more
apparent. Fig.6 shows clearly bimodal energy distribution histograms for $|J_{1}|/|J_{2}|=1.3$, in which
the dip between the peaks is observable already at smaller $L$, quite rapidly approaching zero as $L$ is
increased, indicating discontinuous behaviour of the energy at a rather strong first-order transition.
Although we do not show it here, similar double peaks can also be observed in the histograms of each LRO
parameter.
\newline
\indent The transition remains first order and simultaneous for
dipole, quadrupole and chiralities ordering until fairly small
values of $|J_{1}|/|J_{2}|$. Below $|J_{1}|/|J_{2}| \simeq 0.25$,
however, quadrupoles order separately at temperatures higher than
those for dipole ordering. Thus the phase boundary branches and a
new middle phase of axial quadrupole long-range order (QLRO)
without magnetic dipole ordering opens between the paramagnetic
and DLRO phases. This phase broadens as $|J_{1}|/|J_{2}|$
decreases, since the QLRO branch is little sensitive to the
$|J_{1}|/|J_{2}|$ ratio variation and levels off, while the DLRO
branch turns down approaching the point
($|J_{1}|/|J_{2}|,k_{B}T/|J_{2}|)=(0,0)$. This means that the
ground state is always magnetic as long as there is a finite
dipole exchange interaction. In Fig.7 we present the temperature
variation of the DLRO, QLRO, CHLRO and QCHLRO parameters $m$, $q$,
$\kappa$ and $\kappa^{q}$, respectively, at
$|J_{1}|/|J_{2}|=0.15$. We can see that quadrupoles order before
dipoles, forming a fairly broad region of QLRO without DLRO. On
the other hand, the chirality and quadrupole chirality seem to
order simultaneously with dipoles and quadrupoles, respectively.
The QLRO transition is apparently second order down to
$|J_{1}|/|J_{2}| = 0$ and the critical indices take the values
$\nu_{Q}=0.50\pm 0.03$, $\gamma_{Q}=1.09 \pm 0.08$ at
$|J_{1}|/|J_{2}|=0.15$ (Fig.8) and $\nu_{Q}=0.520 \pm 0.003$ and
$\gamma_{Q}=1.072 \pm 0.009$ at $J_{1}=0$. In the case of
$J_{1}=0$, the QLRO transition temperature is located as
$k_{B}T_{q}/|J_{2}|=0.729 \pm 0.002$. On the other hand, in the
case of the DLRO transition, the first order seems to persist even
after the QLRO and DLRO boundaries separate for a small range of
the exchange ratio values just below the splitting point. This is
clearly seen in Fig.9 from the distribution diagrams of the DLRO
and QLRO parameters. Although at first glance it seems that both
transitions occur at the same temperature and are of first order,
a closer look reveals that while the bimodal distribution of the
DLRO parameter is between the disordered and ordered states, the
bimodal distribution of the QLRO parameter is between two ordered
states of different finite QLRO parameter values. Therefore, here,
the QLRO parameter only shows a discontinuity within the QLRO
region, rather than paramagnetic-QLRO transition. The first-order
DLRO transition changes to the second-order one upon further
lowering of $|J_{1}|/|J_{2}|$. This is seen from the finite-size
scaling analysis of the HMC data for $|J_{1}|/|J_{2}|=0.15$
(Fig.10). The slopes apparently indicate the second-order
character of the transition with the critical indices
$\nu_{M}=0.63\pm 0.02$, $\gamma_{M}=1.25 \pm 0.04$. The resulting
phase diagram is drawn in Fig.11 and some relevant numerical
results listed in Table 1. \vspace{6mm}
\newline
\noindent {\bf\Large{V. Summary and discussion}}\vspace{3mm}
\newline
\indent We studied effects of the biquadratic exchange on the
phase diagram of the frustrated classical XY antiferromagnet on
STL. This study, which to our best knowledge is first for the
studied system, covered most of the significant phenomena induced
by the presence of the biquadratic exchange, and present a fairly
compact picture of the role of this higher-order exchange
interaction on the critical behaviour of the system considered. We
obtained the phase diagram with two ordered phases: in the region
where the bilinear exchange is dominant there is a single phase
transition to the DLRO phase, which is second order at $J_{2} =
0$, but changes to a first-order one upon adding of a rather small
amount of biquadratic exchange. In the region of small
$|J_{1}|/|J_{2}|$ the phase boundary splits into the QLRO
transition line at higher temperatures and the DLRO transition
line at lower temperatures, which are second order, and partly
first and partly second order, respectively.
\newline
\indent From our qualitative and quantitative evaluations we found
out that not only the order of the transitions in different
regions of the $|J_{1}|/|J_{2}|$ parameter is not the same, but
also the sets of the critical indices obtained in different
regions of the second-order transition are different for seemingly
the same kind of transition while almost identical for different
kinds of transition. Let us first discuss the problem of the order
of the transition. The second-order transition at $J_{2} = 0$ is
in agreement with the previous MC studies \cite{kawamura1,diep}
but in contradiction with the renormalization group study
\cite{antonenko}, which predicts a clear first-order transition.
At finite $J_{2}$, the first-order transition observed in the
region of the paramagnetic-DLRO transition has also been observed
in the case of a ferromagnet with $J_{1}>0$, $J_{2}^{\perp}>0$ and
$J_{2}^{\parallel}>0$, however, only in a quite narrow region of
$J_{1}/J_{2}\in(0.33,0.55)$ \cite{zukovic}. We believe that the
mechanism responsible for this transition in the present case is
similar to that in the case of the ferromagnet, i.e., it could
result from a kind of tension between the bilinear and biquadratic
exchange interactions, which in the present case seems to be
enhanced by the presence of the frustration and consequently
causing broadening of the first-order transition region. Namely,
while the decreasing bilinear exchange drives the transition
temperature down to the lower values, the biquadratic exchange
does not follow this tendency and rather prevents the ordering
temperature from rapid decrease. This tendency is clearly seen
from the phase diagram both in the region of separate transitions,
where $T_{q}$ does not vary much with decreasing
$|J_{1}|/|J_{2}|$, as well as in the region of simultaneous
ordering, where the transition temperature is apparently enhanced
by the presence of the biquadratic exchange (the case of absent
biquadratic exchange is represented by the dash-dot straight line
in ($|J_{1}|-k_{B}T_{c}$) parameter space). Put differently,
quadrupoles would prefer ordering at higher temperatures but as
long as there is a single transition they are prevented to do so
by too low bilinear exchange, and order occurs only if the
temperature is lowered still further. This ``frustration" results
in a first-order transition when the strength of the quadrupole
ordering prevails and frustrated quadrupoles order abruptly along
with dipoles. However, when $|J_{2}|$ reaches high values the
frustration becomes too high for the two kinds of ordering to
occur simultaneously and they separate. In order to understand the
first-order DLRO transition and QLRO parameter discontinuity in
the region just below the point of the separation, we analyzed
snapshots (not shown) for $|J_{1}|/|J_{2}|=0.25$ just before the
DLRO sets in. In the snapshots we could observe fairly large
clusters of antiferromagnetically ordered spins along the stacking
direction, which is non-frustrated and in which spins seem to
order more easily than within frustrated planes (Note that in the
case of the non-frustrated parallel (ferromagnetic) ordering the
transition temperature is roughly twice as higher as in the
present case \cite{zukovic}). These clusters reorient at the
transition as a whole, and such a way may produce discontinuities
in the order parameter and internal energy i.e., a first-order
transition. Besides those clusters, we could also observe smaller
intra-plane clusters of spins the axes of which show local
parallel ordering. At the DLRO transition, the spins in these
clusters (and also their axes) reorient into the $120^{\circ}$
spin structure, which may result in the small discontinuity of the
QLRO parameter, seen in Fig.9. The separate QLRO is apparently
second order, in agreement with the mapping arguments of Carmesin
\cite{carmesin}.
\newline
\indent Now, let us address the problem of the critical indices in the case of a second-order
transition. For the case of $J_{2}=0$, there has been an argument about the universality class of the
critical behavior of such a system. Kawamura claimed that it should display a nonstandard (chiral)
universality class behavior due to a two-fold discrete degeneracy $Z_{2}$ associated with the two chiral
states, with the novel indices ($\alpha_{M}=0.34\pm 0.06$, $\beta_{M}=0.253\pm 0.01$, $\gamma_{M}=1.13
\pm 0.05$ and $\nu_{M}=0.54\pm 0.02$) \cite{kawamura1}, while Plumer $et\ al.$ \cite{plumer2} maintained
that there is no new universality class and that the indices take the mean-field tricritical values
($\alpha_{M}=\frac{1}{2}$, $\beta_{M}=\frac{1}{4}$, $\gamma_{M}=1$ and $\nu_{M}=\frac{1}{2}$). As we can
see, the DLRO critical indices obtained from our calculations $\nu_{M}=0.52\pm 0.03$, $\gamma_{M}=1.08
\pm 0.08$ are somewhere between those from Refs.\cite{kawamura1,plumer2} and, considering the error
estimates, could be interpreted for support of either of the theories. Although we can make no definite
conclusion based on the values of the indices themselves, we believe that the former interpretation is
more favorable. Indeed, looking at the critical indices of the separate QLRO transitions we can see that
they are strikingly similar to those for the case of $J_{2}=0$ (and seem to be such along the whole
paramagnetic-QLRO boundary). These indices can hardly be interpreted as the mean-field tricritical ones
and the theory of the same universality class critical behavior of quadrupoles ($J_{1}=0$) and dipoles
($J_{2}=0$), based on mapping and quantitative analysis \cite{zukovic2}, would rather strongly suggest
that both cases show the chiral universality class behavior. As far as the separate DLRO transition is
concerned, in the second-order transition region we obtained the critical indices $\nu_{M}=0.63\pm
0.02$, $\gamma_{M}=1.25 \pm 0.04$, which are quite different from those for the DLRO transition at
$J_{2}=0$. The reason is that in this case the transition has an Ising-like character and, hence, the
indices take on the Ising universality class values ($\nu^{Ising} = 0.629$, $\gamma^{Ising} = 1.239$
\cite{ferr-landau}). The situation is illustrated in Figs.12(a,b). In the QLRO (and no DLRO) region,
there is an axial quadrupole ordering on each of the three sublattices (Fig.12(a)) and only upon further
lowering of the temperature the system reaches the QLRO+DLRO phase in which the Ising-like directional
dipole ordering within the given axis in each sublattice takes place (Fig.12(b)). Therefore, here, the
only difference from the Ising case is that dipoles can order along any of the three axes, not only the
z-axis.
\newline
\indent Our further intention is to perform similar simulations on the STL antiferromagnet for some
other interesting cases, like: $J_{2}^{\perp}<0$, $J_{2}^{\parallel}<0$; $J_{2}^{\perp}>0$,
$J_{2}^{\parallel}>0$; $J_{2}^{\perp}>0$, $J_{2}^{\parallel}<0$. Besides the geometrical frustration,
such spin systems will possess additional frustration arising from the bilinear and biquadratic
exchanges competing in the stacking direction, intra-plane direction, and both stacking and intra-plane
directions, respectively.


\begin{figure}[!t]
\includegraphics[scale=0.5]{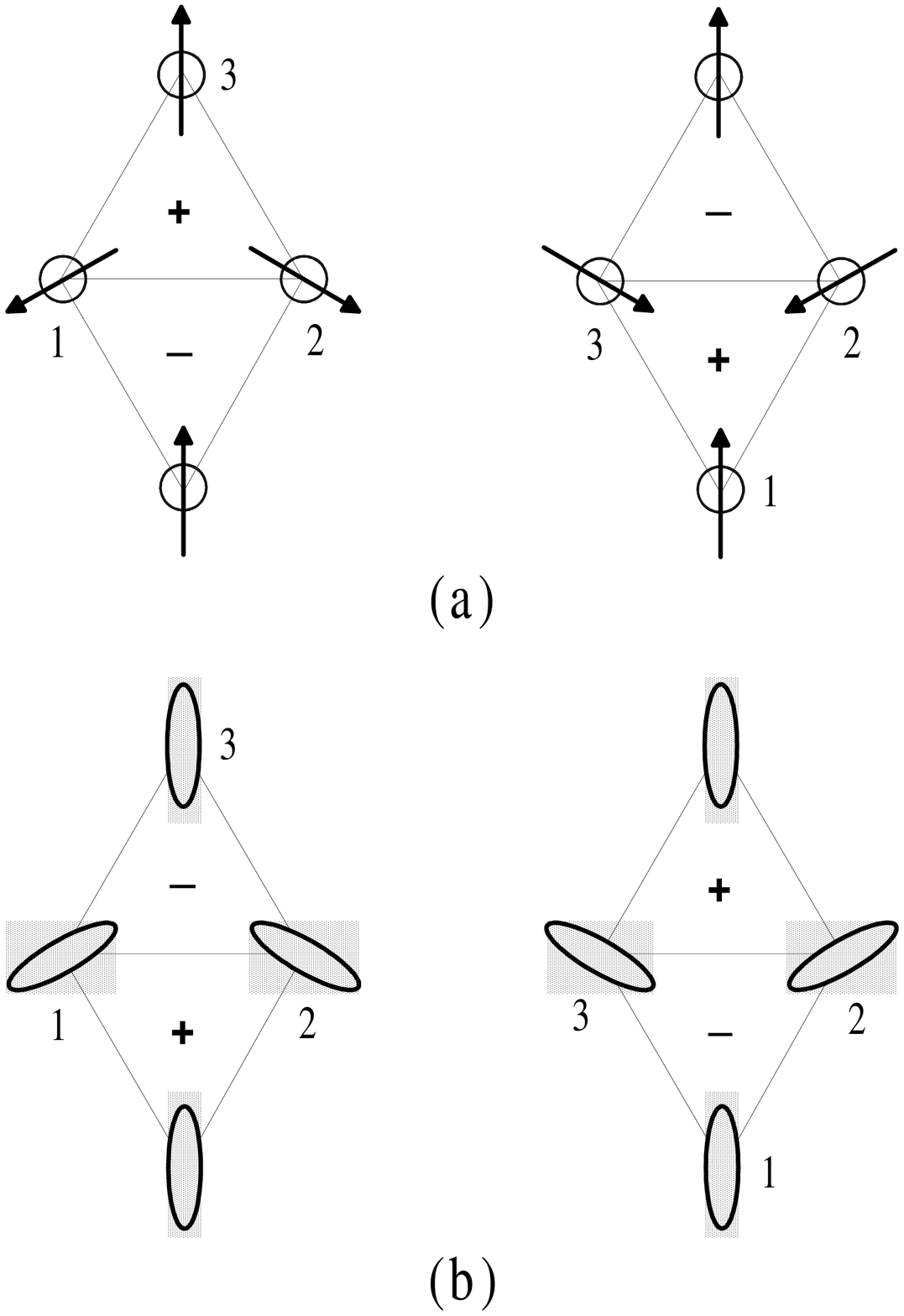}
\caption{Two degenerate ground states, $+120^{\circ}$ and $-120^{\circ}$ structures
on (a) spin and (b) quadrupole plaquettes. Signs $+$ and $-$ denote the sign of (a) chirality and (b)
quadrupole chirality of the elementary triangles. Spins and quadrupoles are numbered counterclockwise,
corresponding to the definitions (\ref{chirality}) and (\ref{q-chirality}).}
\end{figure}

\begin{figure}[!t]
\includegraphics[scale=0.5]{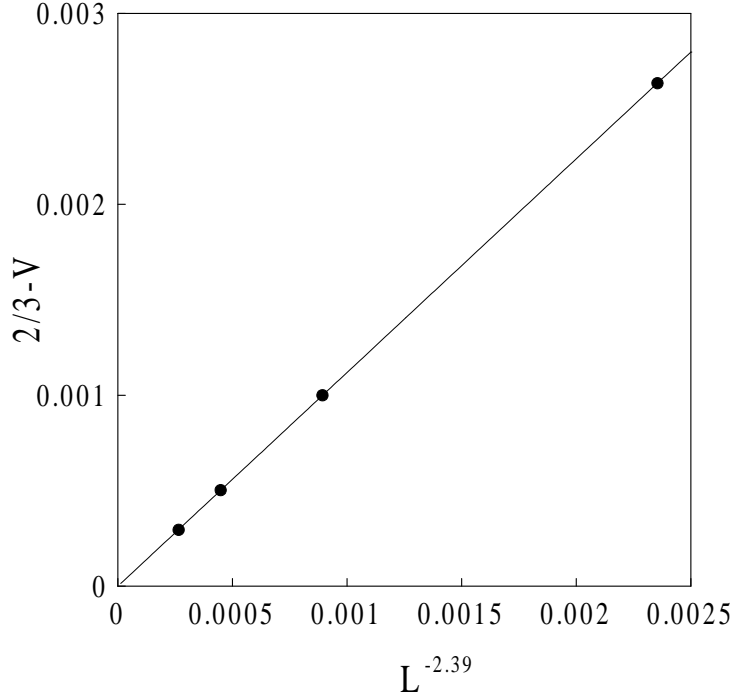}
\caption{Scaling of the energy cumulant minima at $J_{2} = 0$. The values extrapolated to $L
\rightarrow \infty$ approach the value $V^{*}=\frac{2}{3}$ and do not scale with volume, as it should be
in the case of a second-order transition.}
\end{figure}

\begin{figure}[!t]
\includegraphics[scale=0.5]{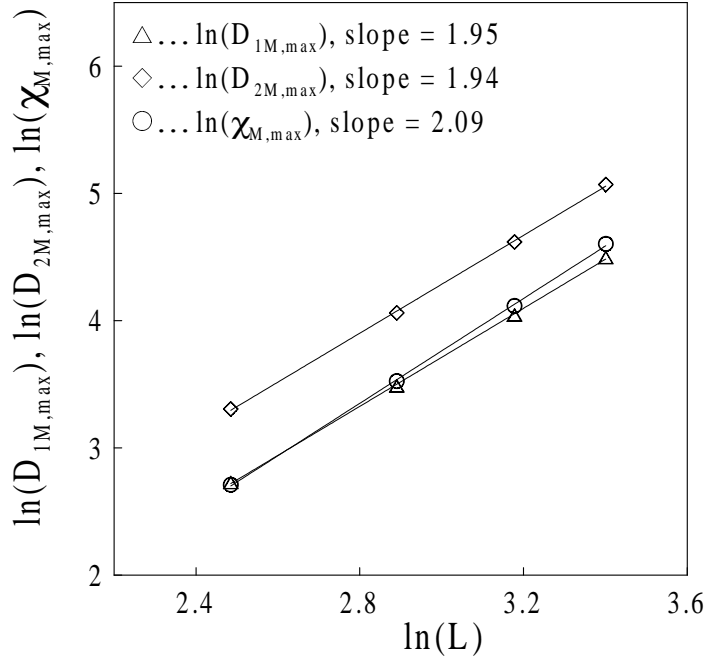}
\caption{Scaling behaviour of the maxima of the susceptibility $\chi_{M,max}$
corresponding to the parameter $M$ and logarithmic derivatives of its first and second moments
$D_{1M,max}$ and $D_{2M,max}$, respectively, in ln-ln plot, for $J_{2} = 0$. The slopes yield values of
$1/\nu_{M}$ for $D_{1M,max},\ D_{2M,max}$ and $\gamma_{M}/\nu_{M}$ for $\chi_{M,max}$.}
\end{figure}

\begin{figure}[!t]
\includegraphics[scale=0.5]{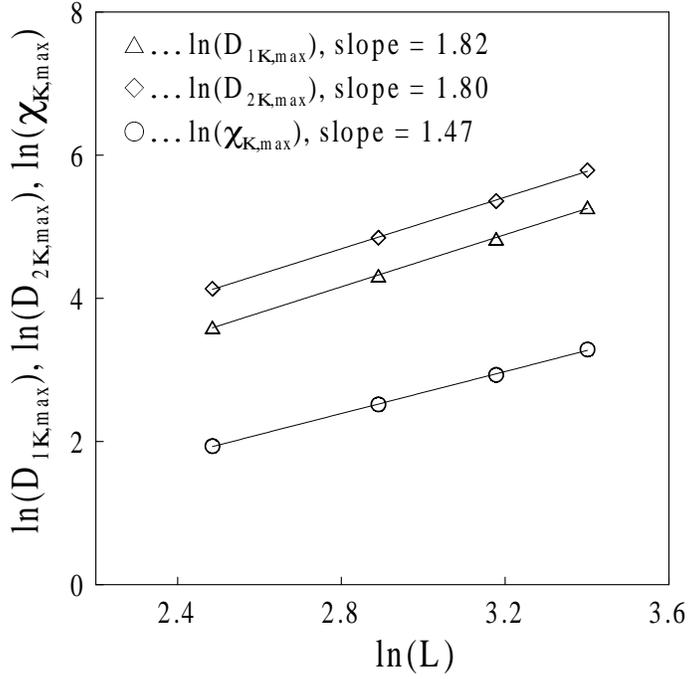}
\caption{The same dependence as in Fig.3, with the parameter $K$ considered instead of $M$.}
\end{figure}

\begin{figure}[!t]
\includegraphics[scale=0.5]{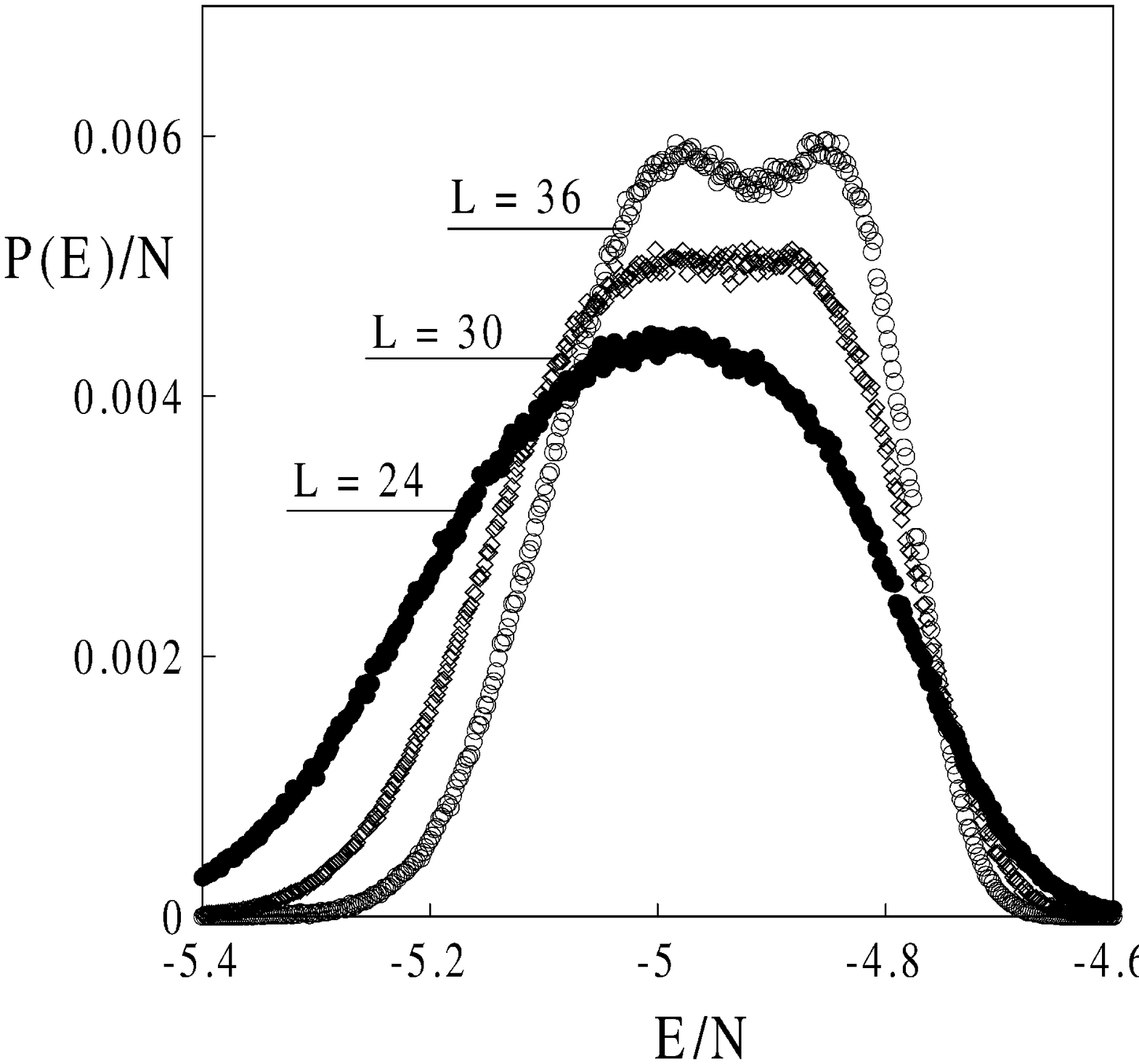}
\caption{Energy distribution at the size-dependent transition temperatures $T_{c}(L)$ for
various lattice sizes and $|J_{2}|/|J_{1}| = \frac{1}{5}$. The bimodal distribution signaling a
first-order transition can only be seen at $L \geq 30$.}
\end{figure}

\begin{figure}[!t]
\includegraphics[scale=0.5]{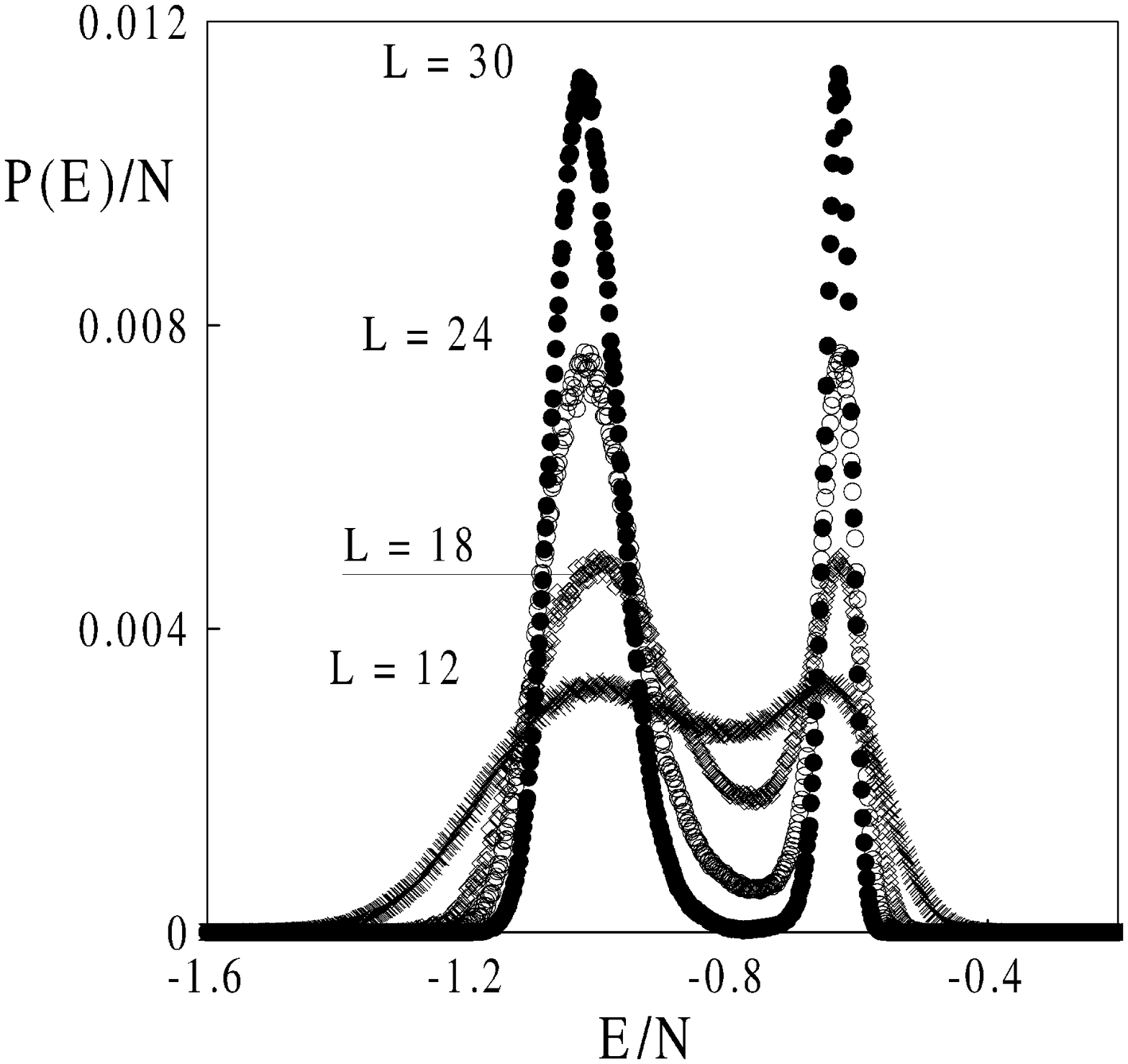}
\caption{Energy distribution at $T_{c}(L)$ for $|J_{1}|/|J_{2}| = 1.3$. Double-peaked
structure with deepening barrier between the two energy states with increasing lattice size indicates a
first-order transition.}
\end{figure}

\begin{figure}[!t]
\includegraphics[scale=0.5]{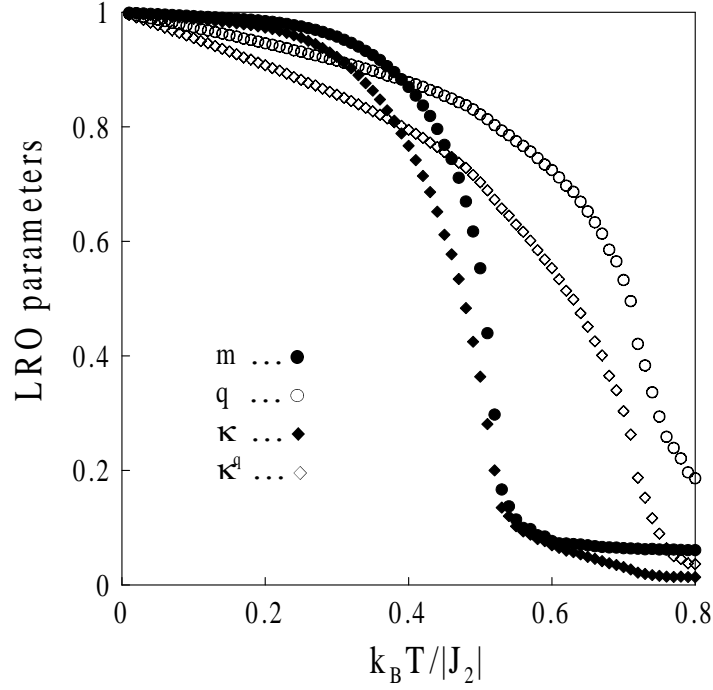}
\caption{Temperature variation of the DLRO, QLRO, CHLRO and QCHLRO parameters $m$, $q$,
$\kappa$ and $\kappa^{q}$, respectively, for $|J_{1}|/|J_{2}| = 0.15$ and $L$ = 12.}
\end{figure}

\begin{figure}[!t]
\includegraphics[scale=0.5]{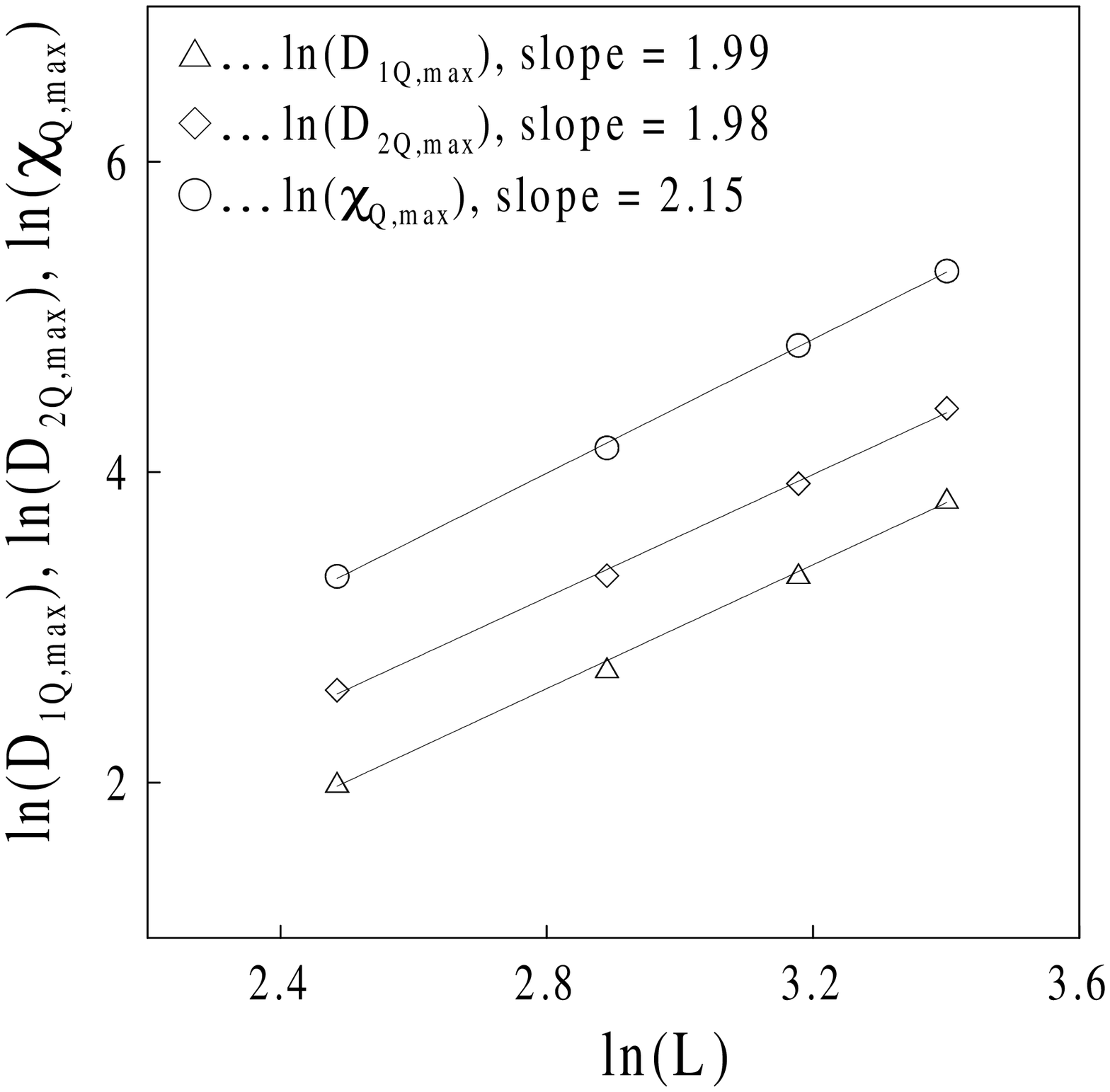}
\caption{Scaling behaviour of the maxima of the susceptibility $\chi_{Q,max}$ and logarithmic
derivatives of the parameter $Q$ and its second moment $D_{1Q,max}$ and $D_{2Q,max}$, respectively, in
ln-ln plot, for $|J_{1}|/|J_{2}| = 0.15$. The slopes yield values of $1/\nu_{Q}$ for $D_{1Q,max},\
D_{2Q,max}$ and $\gamma_{Q}/\nu_{Q}$ for $\chi_{Q,max}$.}
\end{figure}

\begin{figure}[!t]
\includegraphics[scale=0.5]{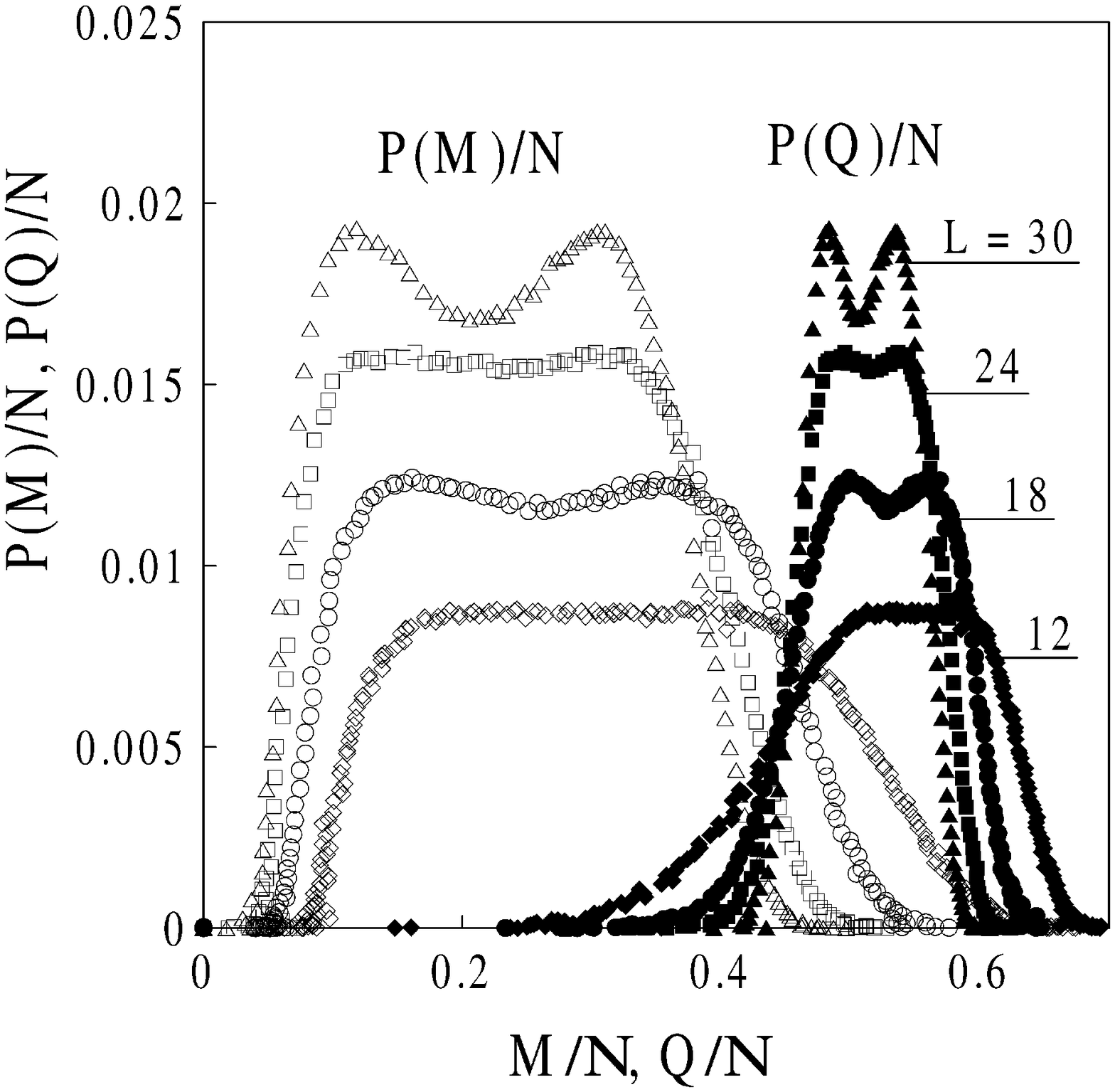}
\caption{Distribution histograms $P(M)$ and $P(Q)$ of DLRO and QLRO parameters, respectively,
at $T_{c}(L)$ for $|J_{1}|/|J_{2}| = 0.25$. The bimodal distributions of the DLRO and QLRO parameters
signal a first-order disorder-DLRO transition and a jump between two finite values of QLRO parameter,
respectively (see text).}
\end{figure}

\begin{figure}[!t]
\includegraphics[scale=0.5]{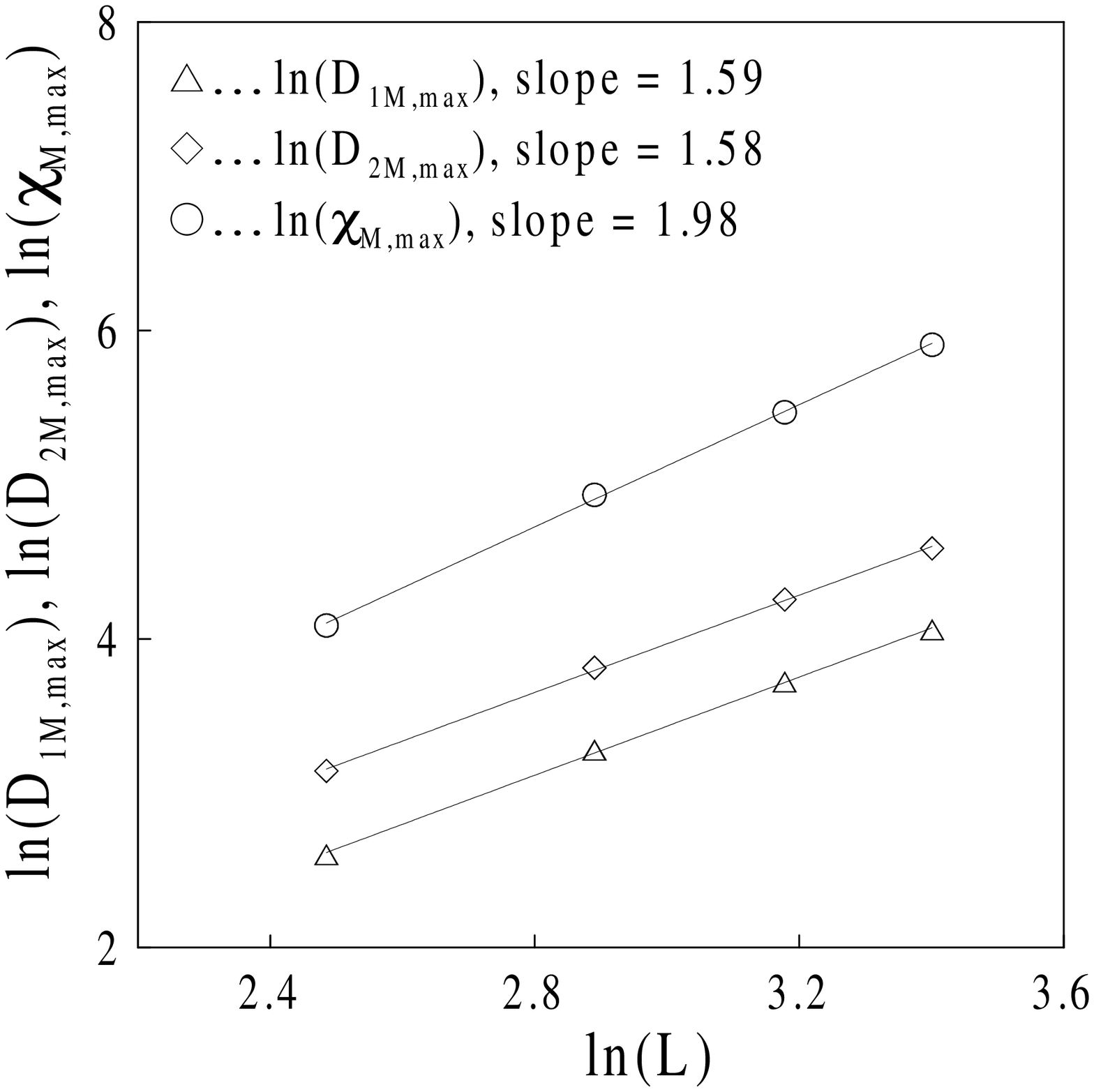}
\caption{Scaling behaviour of the maxima of the susceptibility $\chi_{M,max}$ and
logarithmic derivatives of the DLRO parameter and its second moment $D_{1M,max}$ and $D_{2M,max}$,
respectively, in ln-ln plot, for $|J_{1}|/|J_{2}| = 0.15$. The slopes yield values of $1/\nu_{M}$ for
$D_{1M,max},\ D_{2M,max}$ and $\gamma_{M}/\nu_{M}$ for $\chi_{M,max}$.}
\end{figure}

\begin{figure}[!t]
\includegraphics[scale=0.5]{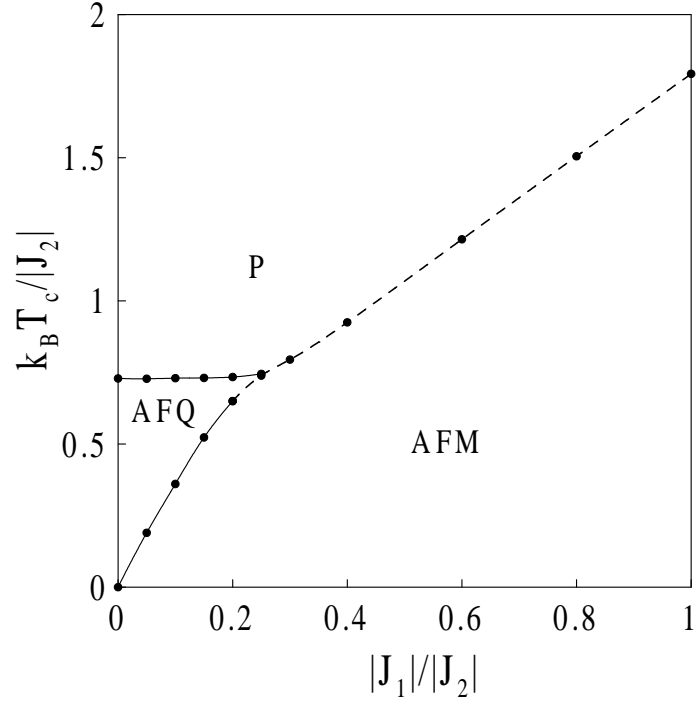}
\caption{Phase diagram in $(|J_{1}|/|J_{2}|,k_{B}T_{c}/|J_{2}|)$ space. The paramagnetic (P), antiferroquadrupolar (AFQ), and antiferromagnetic (AFM) regions correspond to the phases in which both dipoles and quadrupoles are disordered, only quadrupoles are ordered, and both dipoles and quadrupoles are ordered, respectively. The solid and dashed lines correspond to second- and first-order transitions, respectively.}
\end{figure}

\begin{figure}[!t]
\includegraphics[scale=0.5]{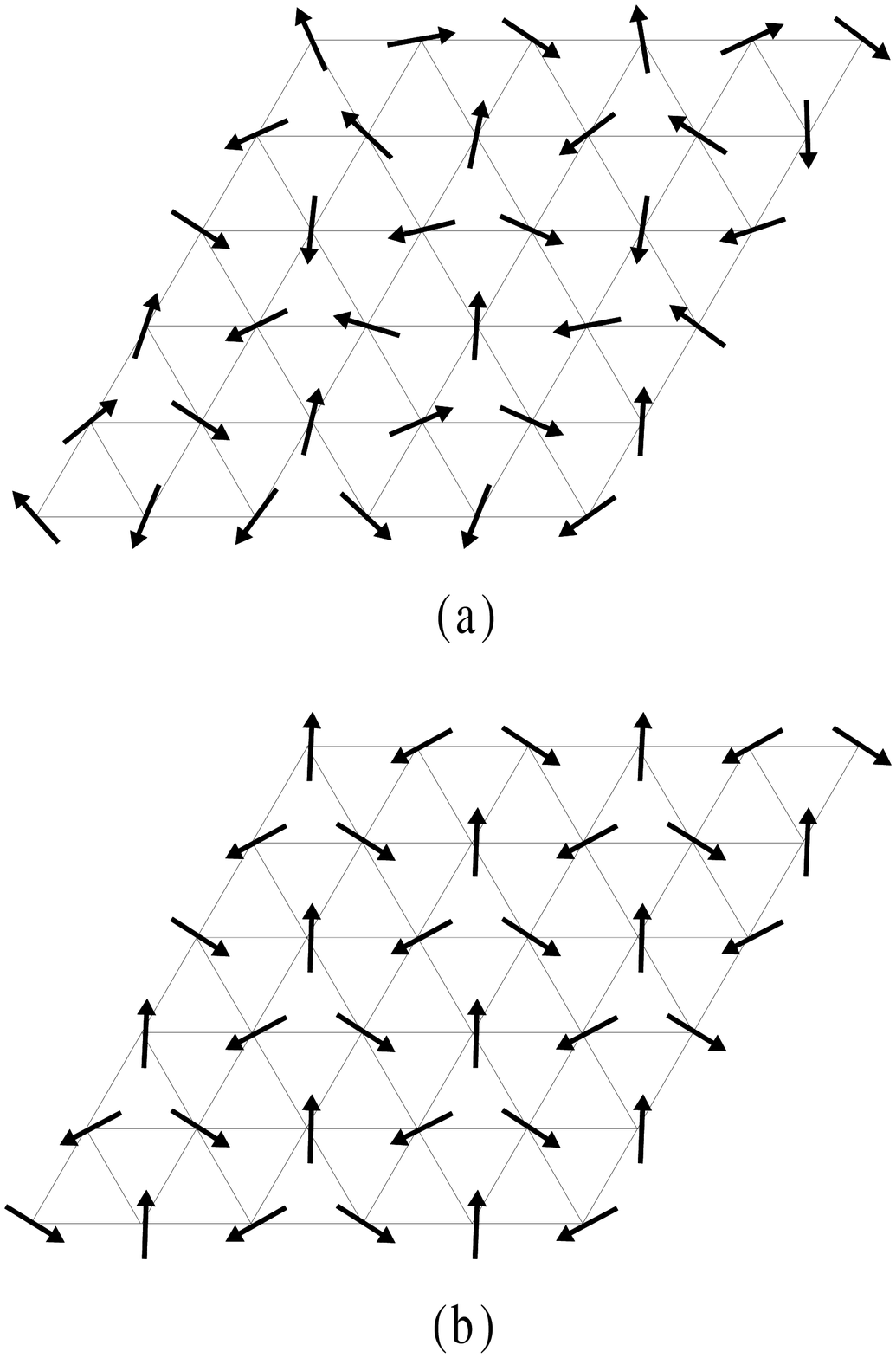}
\caption{Spin configuration snapshots of the system for $|J_{1}|/|J_{2}| = 0.05$ in (a) QLRO
phase ($k_{B}T/|J_{2}|=0.3$) and (b) DLRO phase ($k_{B}T/|J_{2}|=0.001$).}
\end{figure}

\newpage

\begin{table}
\caption{Critical indices and transition temperatures for quadrupole, dipole, and chiral ordering,
respectively.} \label{tab.1}
\begin{center}
\begin{tabular}{|c||c|c|c|}                                                                            \hline
 $|J_{1}|/|J_{2}|$  & $\nu_{Q}$            & $\gamma_{Q}$           & $k_{B}T_{q}$                  \\ \hline
 0                  & 0.520 $\pm$ 0.003    & 1.072 $\pm$ 0.009      & 0.729 $\pm$ 0.002 $|J_{2}|$   \\ \hline
 0.15               & 0.50 $\pm$ 0.03      & 1.09 $\pm$ 0.08        & 0.731 $\pm$ 0.001 $|J_{2}|$   \\ \hline \hline
                    & $\nu_{M}$            & $\gamma_{M}$           & $k_{B}T_{c}$                  \\ \hline
 0.15               & 0.63 $\pm$ 0.02      & 1.25 $\pm$ 0.04        & 0.523 $\pm$ 0.002 $|J_{2}|$   \\ \hline
 $\infty$           & 0.52 $\pm$ 0.03      & 1.08 $\pm$ 0.08        & 1.4580 $\pm$ 0.0005 $|J_{1}|$ \\ \hline \hline
                    & $\nu_{\kappa}$       & $\gamma_{\kappa}$      & $k_{B}T_{\kappa}$             \\ \hline
 $\infty$           & 0.55 $\pm$ 0.01      & 0.81 $\pm$ 0.03        & 1.4590 $\pm$ 0.0013 $|J_{1}|$ \\ \hline

\end{tabular}
\end{center}
\end{table}


\newpage

\begin{thebibliography}{}

\bibitem{sivar72} J. Sivardiere,
        {\em Phys. Rev. B} {\bf 6}, 4284, (1972).
\bibitem{sivar73} J. Sivardiere, A.N. Berker and M. Wortis,
        {\em Phys. Rev. B} {\bf 7}, 343, (1973).
\bibitem{chen-levy1} H.H. Chen and P. Levy,
        {\em Phys. Rev. Lett.} {\bf 27}, 1383, (1971); {\em Phys. Rev. B} {\bf 7}, 4267, (1973).
\bibitem{chen-levy2} H.H. Chen and P. Levy,
        {\em Phys. Rev. B} {\bf 7}, 4284 , (1973).
\bibitem{micnas} R. Micnas,
        {\em J. Phys. C: Solid St. Phys.} {\bf 9}, 3307, (1976).
\bibitem{chaddha99} G.S. Chaddha and A. Sharma,
        {\em J. Magn. Magn. Mater.} {\bf 191}, 373, (1999).
\bibitem{chen-etal1} K.G. Chen, H.H. Chen, C.S. Hsue and F.Y. Wu,
        {\em Physica} {\bf 87A}, 629, (1977).
\bibitem{chen-etal2} K.G. Chen, H.H. Chen and C.S. Hsue,
        {\em Physica} {\bf 93A}, 526, (1978).
\bibitem{tanaka} A. Tanaka and T. Idogaki,
        {\em J. Phys. Soc. Japan} {\bf 67}, 604, (1998).
\bibitem{campbell} M. Campbell and L. Chayes,
        {\em J. Phys. A} {\bf 32}, 8881, (1999).
\bibitem{nagata} H. Nagata, M.\v{Z}ukovi\v{c} and T.Idogaki,
        {\em J. Magn. Magn. Mater.} {\bf 234}, 320, (2001).
\bibitem{zukovic} M.\v{Z}ukovi\v{c}, T.Idogaki and K.Takeda,
        {\em Physica B} {\bf 304}, 18, (2001).
\bibitem{kawamura1} H. Kawamura, {\em{J. Phys. Soc. Japan}} {\bf 54}, 3220, (1985); {\bf 55}, 2095, (1986);
        {\bf 56}, 474, (1986); {\bf 58}, 584, (1989); {\bf 61}, 1299, (1992).
\bibitem{kawamura2} H. Kawamura, {\em{Phys. Rev.}} {\bf B38}, 4916, (1988); {\bf B42}, 2610 (E), (1990);
        {\em{J. Phys. Soc. Japan}} {\bf 59}, 2305, (1990).
\bibitem{plumer1} M.L. Plumer, A. Mailhot and A. Caill\'{e},
        {\em Phys. Rev. B} {\bf 48}, 3840, (1993).
\bibitem{diep} E.H. Boubcheur, D. Loison and H.T. Diep,
        {\em Phys. Rev. B} {\bf 54}, 4165, (1996).
\bibitem{ferr-swen1} A.M. Ferrenberg and R.H. Swendsen,
        {\em Phys. Rev. Lett.} {\bf 61}, 2635 , (1988).
\bibitem{ferr-swen2} A.M. Ferrenberg and R.H. Swendsen,
        {\em Phys. Rev. Lett.} {\bf 63}, 1195 , (1989).
\bibitem{miyashita} S. Miyashita and H. Shiba,
        {\em{J. Phys. Soc. Japan}} {\bf 53}, 1145, (1984).
\bibitem{antonenko} A. Antonenko and A. I. Sokolov,
        {\em Phys. Rev. B} {\bf 49}, 15901, (1994).
\bibitem{carmesin} H.-O. Carmesin,
        {\em Phys. Lett. A} {\bf 125}, 294, (1987).
\bibitem{plumer2} M.L. Plumer and A. Mailhot,
        {\em Phys. Rev. B} {\bf 50}, 16113, (1994).
\bibitem{zukovic2} M.\v{Z}ukovi\v{c}, T.Idogaki and K.Takeda,
        {\em Phys. Rev. B} {\bf 63}, 172412-1, (2001).
\bibitem{ferr-landau} A.M. Ferrenberg and D.P. Landau,
        {\em Phys. Rev. B} {\bf 44}, 5081, (1991).

\end{thebibliography}
\end{document}